\documentclass[12pt]{article}

\usepackage[margin=1in]{geometry}
\usepackage{amsmath, amssymb}
\usepackage{booktabs}
\usepackage{graphicx}
\usepackage{natbib}
\usepackage{setspace}
\usepackage{hyperref}
\usepackage{tikz}
\usepackage{pgfplots}
\pgfplotsset{compat=1.18}

\title{The Economics of War: Militarization and Growth in an AK Economy}
\author{
Arpan Chakraborty\thanks{PhD Scholar, Department of Humanities and Social Sciences, IIT Kharagpur.}
}
\date{\today}

\begin{document}

\maketitle

\begin{abstract}
This paper analyzes the macroeconomic consequences of military spending and militarization within a dynamic growth framework. Building on a Keynesian goods-market model, we examine how the allocation of government expenditure between civilian and military sectors affects capital accumulation and technological progress. Military spending generates opposing effects: it stimulates aggregate demand and may support innovation through defense-related research, but it also crowds out civilian investment and creates structural rigidities. We formalize these mechanisms in a stylized endogenous-growth model in which productivity depends on the degree of militarization, producing a non-linear relationship between the military burden and long-run growth. Calibrated simulations show that moderate levels of military spending can temporarily support growth, whereas excessive militarization reduces long-run development. We further illustrate the asymmetric growth costs of conflict using a simple two-country war simulation between an advanced economy and a sanctioned middle-income economy.
\end{abstract}

\noindent\textbf{Keywords:} Militarization; Economic growth; AK model; Two country model.

\noindent\textbf{JEL Classification:} H56, E12, O41, F51.
% \end{abstract}

\section{Introduction}

The economic implications of war and militarism have been widely debated in political
economy. Governments allocate substantial financial resources to defense budgets, arms
procurement, and military infrastructure. Such expenditures are often justified on the
grounds of national security and geopolitical competition. However, economists have long
questioned whether military spending promotes sustainable economic growth or diverts
resources from socially productive sectors such as education, health, and green
infrastructure.

Historically, war economies have played a central role in industrial development.
Large-scale military mobilization during major conflicts has frequently accelerated
technological innovation and industrial expansion. At the same time, the persistence of
high military expenditure during peacetime has led scholars to describe modern economies
as operating within a \emph{permanent war economy}. From a macroeconomic perspective,
this raises the question of whether the institutionalization of defense production
supports or hinders long-run development.

This paper offers a theory-based contribution to this debate. Rather than focusing on
reduced-form regressions, it builds a sequence of analytical models that explicitly
trace the channels through which militarism affects aggregate demand, capital
accumulation, innovation, and structural dynamics. The starting point is a standard
Keynesian income-expenditure identity with military spending as one component of
government demand. We then extend this setup into a dynamic endogenous-growth model of
the AK type in which military spending influences both the level and the growth rate of
output through its effects on investment and the productivity of capital. The AK
assumption allows us to keep the analysis transparent while foregrounding the role of
the composition of demand in shaping long-run growth, in the spirit of
\citet{Barro1990} and related calibrated macroeconomic models.

Methodologically, the paper combines formal modeling with calibrated simulations. Using
parameter values loosely inspired by the literature on savings, depreciation, and
productivity-enhancing public spending, we derive a non-linear growth function in the
military burden and illustrate the implied trade-offs graphically. We also extend the
model to a two-country environment to study the macroeconomic implications of an
interstate war between an advanced economy and a middle-income economy under sanctions.
The aim is not to estimate structural parameters for particular countries, but to
provide a transparent theoretical framework within which different regimes of
militarization and conflict can be compared.

The rest of the paper is organized as follows. Section 2 presents the literature review. Section 3 discusses the short-run aggregate demand analysis. Section 4 introduces the AK model. Section 5 describes the model parameters. Section 6 presents the simulation results. Section 7 extends the model to a two-country scenario. Finally, Section 8 concludes.

\section{Literature and Conceptual Framework}

The literature on military expenditure and economic performance spans multiple
traditions. Keynesian analyses emphasize the demand-stimulating role of government
spending. Defense budgets, in this view, can raise output and employment when economies
are demand-constrained, operating through multipliers that depend on consumption
behavior and investment responsiveness \citep{Keynes1936}. In such settings, the
composition of government expenditure may matter less in the short run than the overall
level, provided that leakages and import propensities are not excessive.

Neoclassical and endogenous-growth approaches stress opportunity costs and the role of
productive public spending. In models with capital accumulation and technology-driven
growth, resources devoted to unproductive government consumption can depress the
long-run growth rate by reducing the share of output available for private investment
and productivity-enhancing public services \citep{Barro1990}. To the extent that
military outlays generate limited direct productive services for the private economy,
they function as a tax on growth, even if they are financed through lump-sum taxation.
Critical surveys such as \citet{Dunne2005} find that empirical results on the
military--growth nexus are often fragile and context-dependent, which reinforces the
case for explicit theoretical modeling of the underlying channels.

Political economy perspectives place militarism within the institutional structure of
capitalism. The formation of a military--industrial complex, the persistence of
defense-oriented coalitions, and the embedding of war-related production in long-term
accumulation paths are central themes of this tradition
\citep{BaranSweezy1966, Melman1970}. Here, the key issue is not only the level of
military spending but also the way it shapes technological trajectories, sectoral
patterns, and income distribution.

The conceptual framework of this paper integrates these strands by distinguishing
between three main channels. First, military spending is a component of aggregate
demand in the short run. Second, the division of output between military and civilian
uses affects the fraction of income that is reinvested in productive capital in the
long run. Third, military spending can influence the rate and direction of
technological change, both positively through defense-related innovation and negatively
through the crowding out of civilian research and distortions in the allocation of
scientific resources.

\section{Short-Run Aggregate Demand and Military Expenditure}

This section presents simple formal models that operationalize the core mechanisms
discussed above. The goal is not to provide definitive structural estimates but to make
explicit the channels through which militarism enters macroeconomic and innovation
dynamics.

\subsection{Aggregate Demand and Military Expenditure}

Consider a closed-economy income identity:
\begin{equation}
  Y = C + I + G_c + G_m,
\end{equation}
where \(Y\) is real output, \(C\) private consumption, \(I\) private investment, \(G_c\)
civilian government spending, and \(G_m\) military spending.

Assume a simple linear consumption function:
\begin{equation}
  C = c_0 + c_1 (Y - T), \quad 0 < c_1 < 1,
\end{equation}
with lump-sum taxes \(T = \tau Y\), \(0 < \tau < 1\). Private investment is assumed to
depend positively on output and negatively on the real interest rate \(r\):
\begin{equation}
  I = i_0 + i_1 Y - i_2 r, \quad i_1 > 0, \; i_2 > 0.
\end{equation}

Solving for equilibrium output yields:
\begin{equation}
  Y = \frac{c_0 - c_1 \tau Y + i_0 - i_2 r + G_c + G_m}{1 - c_1 - i_1}.
\end{equation}
Rearranging gives:
\begin{equation}
  Y = \frac{c_0 + i_0 - i_2 r + G_c + G_m}{1 - c_1 (1-\tau) - i_1} \equiv k_0 + k_1 G_m,
\end{equation}
where the short-run multiplier of military expenditure is
\begin{equation}
  \frac{\partial Y}{\partial G_m} = \frac{1}{1 - c_1 (1-\tau) - i_1} > 0
\end{equation}
under standard stability conditions. From a purely Keynesian perspective, higher
military spending raises output and employment in the short run. This provides a clear
rationale for using defense budgets as countercyclical instruments in recessions or
periods of insufficient aggregate demand.

However, the same structure already contains the seeds of longer-run tensions. For a
given tax rate \(\tau\), higher \(G_m\) either implies lower \(G_c\) or higher deficits
and future taxation. Moreover, the total level of output determined by this system also
affects the resources available for capital accumulation and technological efforts. The
next section extends the analysis by introducing capital dynamics and a growth
mechanism that depends on the composition of spending.

\section{Long-run Dynamic Theory of Militarized Growth}

\subsection{Capital Accumulation in an AK Framework}

To examine the long-run consequences of militarism, consider a one-sector growth model
with capital accumulation of the AK type\footnote{The AK framework is chosen for analytical clarity. Its linearity in capital allows us to isolate the effects of the composition of government spending without introducing diminishing returns or transitional dynamics. In this setting, changes in the effective investment share or productivity translate directly into changes in the long-run growth rate. The specification is closely related to endogenous-growth models in which public expenditure enters multiplicatively with private capital \citep{Barro1990}. In contrast, in a Solow-type model with exogenous technical progress, militarization would mainly affect the steady-state level of income rather than the long-run growth rate. Semi-endogenous growth models in which research effort drives innovation would yield qualitatively similar implications to our AK formulation.}. Let \(K_t\) denote the stock of productive
capital at time \(t\), and assume that output is given by
\begin{equation}
  Y_t = A_t K_t,
\end{equation}
where \(A_t\) is a productivity term that may depend on the history of spending, and
\(K_t\) represents the reproducible capital stock used in the civilian economy. The AK
assumption implies a constant marginal product of capital at a given \(A_t\), which is
a standard device in endogenous-growth theory.

Suppose that a constant fraction \(s\) of \emph{civilian} income is invested in the
capital stock each period, whereas military spending is treated as current expenditure
that does not directly add to the private capital stock. If \(m_t\) denotes the
\emph{military burden}, defined as the ratio of military spending to output,
\begin{equation}
  m_t = \frac{G_{m,t}}{Y_t},
\end{equation}
and if we abstract from deficits so that total government spending is financed out of
taxes on current output, then the share of output available for civilian uses is
approximately \(1 - m_t\). Investment in the productive capital stock is therefore
\begin{equation}
  I_t^c = s (1 - m_t) Y_t.
\end{equation}
Capital evolves according to
\begin{equation}
  K_{t+1} = (1 - \delta) K_t + I_t^c = (1 - \delta) K_t + s (1 - m_t) A_t K_t,
\end{equation}
where \(0 < \delta < 1\) is the depreciation rate. The gross growth factor of capital
is
\begin{equation}
  \frac{K_{t+1}}{K_t} = 1 - \delta + s (1 - m_t) A_t.
\end{equation}
In a balanced-growth environment with constant \(A_t = A\) and constant military burden
\(m_t = m\), the long-run growth rate of capital and output is
\begin{equation}
  g(m) \equiv \frac{K_{t+1} - K_t}{K_t} = -\delta + s (1 - m) A.
  \label{eq:gm_linear}
\end{equation}
In this baseline specification, military spending unambiguously reduces the long-run
growth rate, because it lowers the fraction of output that is reinvested in civilian
capital. The linear dependence of \(g(m)\) on \(m\) in \eqref{eq:gm_linear} captures
the pure crowding-out channel emphasized in neoclassical analyses of government
consumption.

\subsection{Military-Induced Innovation and a Non-Linear Growth Function}

The historical role of defense spending in generating technological advances suggests
that the assumption of constant \(A\) is too restrictive. Let us therefore model
productivity as a function of the military burden. A simple reduced-form representation
of defense-induced innovation is
\begin{equation}
  A(m) = A_0 \bigl[1 + \varphi m - \chi m^2\bigr],
  \label{eq:Aofm}
\end{equation}
where \(A_0 > 0\) is baseline productivity in the absence of militarism,
\(\varphi > 0\) captures the positive effect of moderate military spending on
technology (for example through defense R\&D and learning-by-doing in high-tech
industries), and \(\chi > 0\) reflects the diminishing and eventually negative returns
of excessive militarization (for example through secrecy, misallocation, and
over-specialization). The quadratic term ensures that the technological contribution of
military spending is bounded and that, beyond some point, further militarization may
actually reduce overall productivity by distorting research priorities and diverting
scientific talent away from socially useful innovation.

Substituting \eqref{eq:Aofm} into \eqref{eq:gm_linear}, we obtain a non-linear growth
function in the military burden:
\begin{equation}
  g(m) = -\delta + s (1 - m) A_0 \bigl[1 + \varphi m - \chi m^2\bigr].
  \label{eq:g_of_m_full}
\end{equation}
This expression formalizes the central theoretical idea of the paper. The term
\(-\delta\) is the loss due to depreciation. The second term reflects the effective
investment effort, which is the product of the civilian investment share \(s(1 - m)\)
and the productivity of capital \(A(m)\). Military spending thus enters the growth rate
through two opposing channels: it directly reduces the share of output available for
civilian investment via the factor \(1 - m\), but it can also raise the productivity of
capital for small to moderate values of \(m\) through the term \(1 + \varphi m - \chi
m^2\). The net effect is theoretically ambiguous and depends on the relative magnitude
of \(\varphi\) and \(\chi\) as well as the baseline parameters \(s\), \(\delta\), and
\(A_0\).

When the positive innovation effect dominates at low levels of militarization but the
crowding-out and distortion effects dominate at high levels, the growth function
\(g(m)\) takes a hump-shaped form. This provides a theoretical justification for the
empirical finding, often reported in the literature, that the relationship between
military burden and growth is weakly non-linear, with some evidence of an interior
maximum at relatively low levels of militarization \citep[see][]{Dunne2005}.

\section{Calibration and Parameter Choices}

To give the model operational content, we calibrate the key parameters in
\eqref{eq:g_of_m_full} using values inspired by the macroeconomic literature. Table
\ref{tab:parameters} summarizes the baseline parameters used in the single-economy
simulations and indicates typical ranges found in the literature.
\begin{table}[h!]
\centering
\small
\caption{Baseline parameter values and links to the literature}
\label{tab:parameters}
\begin{tabular}{@{}llp{3.2cm}p{5.5cm}@{}}
\toprule
Parameter & Baseline value & Interpretation & Literature motivation \\ \midrule
\(s\) & 0.20 & Savings rate out of civilian income &
Typical calibrated value in macro models; see \citet{Barro1990} and RBC calibrations. \\[0.3em]

\(\delta\) & 0.05 & Capital depreciation rate &
Annual depreciation around 5\% is standard in growth and business-cycle models. \\[0.3em]

\(A_0\) & 0.30 & Baseline productivity of capital &
Chosen so that, with \(s\) and \(\delta\), the non-militarized economy has modest growth. \\[0.3em]

\(\varphi\) & 5.0 & Positive innovation effect of \(m\) &
Captures empirical evidence that military R\&D can support high-tech sectors. \\[0.3em]

\(\chi\) & 60.0 & Negative curvature of \(A(m)\) &
Ensures diminishing and eventually negative returns to militarization; consistent with political economy critiques \citep{Melman1970}. \\ 
\bottomrule
\end{tabular}
\end{table}
The savings rate \(s = 0.2\) and depreciation rate \(\delta = 0.05\) are in line with
standard macroeconomic calibrations used in endogenous-growth and real business-cycle
models. The baseline productivity parameter \(A_0 = 0.3\) is chosen so that the
non-militarized economy exhibits a modest positive growth rate. The innovation
parameters \(\varphi\) and \(\chi\) are selected to reflect the idea that moderate
military spending can have appreciable technological benefits but that these benefits
are quickly offset as militarization becomes excessive. Their precise values are not
empirically estimated; rather, they are set to produce a qualitatively plausible
hump-shaped growth function, consistent with the mixed evidence summarized in
\citet{Dunne2005}.

With these parameter values, the growth function becomes
\begin{equation}
  g(m) = -0.05 + 0.2 (1 - m) \, 0.3 \bigl[1 + 5 m - 60 m^2\bigr],
  \label{eq:g_calibrated}
\end{equation}
which can be evaluated for different values of the military burden \(m\) in the range
from 0 to, say, 8\% of GDP. The resulting function is used to simulate the impact of
alternative militarization regimes on the long-run growth rate of the economy.

\section{Simulation Results for a Single Economy}

Evaluating the calibrated growth function \eqref{eq:g_calibrated} reveals three
qualitatively distinct regimes. For very low levels of militarization, the growth rate
is positive but modest, reflecting limited defense-induced innovation and relatively
high civilian investment. As the military burden increases from zero, the positive
effect of military-induced innovation more than compensates for the loss of civilian
investment share. This pushes up the growth rate, which reaches a local maximum at an
intermediate military share. Beyond this point, further increases in the military
burden reduce growth. The decline is driven by both the shrinking civilian investment
share and the negative curvature of the productivity function as excessive militarism
distorts technological trajectories.

\begin{figure}[h!]
  \centering
  \begin{tikzpicture}
    \begin{axis}[
      width=0.8\textwidth,
      height=0.45\textwidth,
      xlabel={Military burden \(m\) (share of GDP)},
      ylabel={Long-run growth rate \(g(m)\)},
      xmin=0, xmax=0.08,
      ymin=-0.05, ymax=0.05,
      grid=both,
      grid style={dotted,gray!40},
      thick,
      axis line style={very thick},
      tick style={very thick},
    ]
      % Calibrated growth function:
      % g(m) = -0.05 + 0.2*(1 - m)*0.3*(1 + 5 m - 60 m^2)
      \addplot[
        blue!75!black,
        domain=0:0.08,
        samples=400,
        very thick
      ]
      { -0.05 + 0.2*(1 - x)*0.3*(1 + 5*x - 60*x^2) };
    \end{axis}
  \end{tikzpicture}
  \caption{Calibrated relationship between the military burden and the long-run growth
  rate. Note: theoretical simulation based on Equation~\eqref{eq:g_calibrated}.}
  \label{fig:g_of_m}
\end{figure}
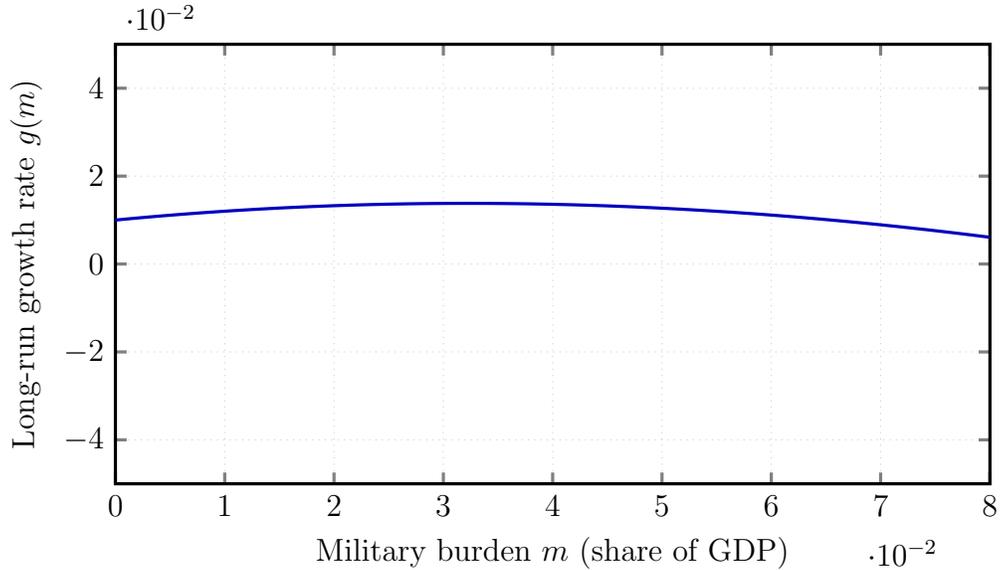

Figure~\ref{fig:g_of_m} depicts the simulated relationship between the military burden
and the long-run growth rate implied by \eqref{eq:g_calibrated}. The hump-shaped curve
illustrates the existence of an interior growth-maximizing level of militarization. At
very low military burdens, potential security and innovation benefits are underutilized
in this stylized framework. At very high burdens, the economy is effectively locked
into a permanent war economy with insufficient civilian accumulation and distorted
technology, leading to secular stagnation or even decline.

The theoretical structure of the model allows for comparative-static insights. An
increase in the savings rate \(s\) raises the overall level of the growth function and
can also shift the location of the interior maximum, potentially making the economy
more tolerant of military burdens before growth begins to decline. A higher
depreciation rate \(\delta\) uniformly lowers growth for all values of \(m\). Changes in
the innovation parameters \(\varphi\) and \(\chi\) alter the curvature of the
productivity function. A larger \(\varphi\) intensifies the positive impact of moderate
militarization, while a larger \(\chi\) steepens the decline at high levels of military
spending, making the economy more vulnerable to over-militarization.

Within this framework, the permanent war economy corresponds to a policy choice or
institutional equilibrium in which the military burden is persistently above the
growth-maximizing level. The resulting regime may be associated with a stable coalition
of actors who benefit from high defense budgets despite the fact that overall growth is
lower than it could be.

\section{A Two-Country War Model}

\subsection{Theoretical Structure}

We now extend the model to a simple two-country environment to study the macroeconomic
implications of an interstate war. Consider two countries, indexed by \(i = U, I\), to
be interpreted as an advanced economy (benchmarking the United States) and a
middle-income economy under sanctions (benchmarking Iran). Each country has an AK-type
technology,
\begin{equation}
  Y_{i,t} = A_i(m_{i,t}) K_{i,t},
\end{equation}
with country-specific productivity functions
\begin{equation}
  A_i(m_i) = A_{0,i} \bigl[1 + \varphi_i m_i - \chi_i m_i^2\bigr], \quad i \in \{U,I\}.
\end{equation}
Capital in each country evolves according to
\begin{equation}
  K_{i,t+1} = (1 - \delta_i - d_{i,t}) K_{i,t} + s_i (1 - m_{i,t}) Y_{i,t},
  \label{eq:Ki_dynamics}
\end{equation}
where \(s_i\) is the savings rate out of civilian income, \(\delta_i\) is the normal
depreciation rate, and \(d_{i,t}\) is an additional destruction rate due to war
(damage to infrastructure, capital loss, disruptions). The term \(m_{i,t}\) is the
military burden in country \(i\). In peacetime, we set \(d_{i,t} = 0\) and assume that
each country chooses a time-invariant military burden \(m_i\). In wartime, both
\(m_{i,t}\) and \(d_{i,t}\) increase, reflecting rapid militarization and physical
destruction. We abstract from strategic interactions and focus on the macroeconomic
consequences of exogenous war shocks.

For each country, the instantaneous growth rate of capital and output is
\begin{equation}
  g_i(m_i, d_i) = -\delta_i - d_i + s_i (1 - m_i) A_{0,i} \bigl[1 + \varphi_i m_i - \chi_i m_i^2\bigr],
  \label{eq:gi}
\end{equation}
where, in peacetime, \(d_i = 0\), and in wartime, \(d_i > 0\) and \(m_i\) jumps upward.

\subsection{Calibration: United States and Iran}

To capture stylized differences between the United States and Iran, we choose
country-specific parameters consistent with macroeconomic magnitudes and qualitative
features discussed in the literature on military spending and growth
\citep[e.g.][]{Dunne2005}. Table~\ref{tab:parameters_two_country} summarizes the
calibration \footnote{The parameters should be updated to reflect data availability in the current wartime context. Future research can incorporate media-based data and rerun the simulations to assess projected long-run growth.}.

\begin{table}[h!]
\centering
\caption{Country-specific parameters for the two-country war model}
\label{tab:parameters_two_country}
\begin{tabular}{@{}llll@{}}
\toprule
Parameter & United States (\(U\)) & Iran (\(I\)) & Interpretation \\ \midrule
\(s_i\) & 0.22 & 0.18 & Savings rate out of civilian income. \\
\(\delta_i\) & 0.05 & 0.06 & Normal depreciation rate. \\
\(A_{0,i}\) & 0.35 & 0.25 & Baseline productivity of capital. \\
\(\varphi_i\) & 6.0 & 4.0 & Positive innovation effect of \(m_i\). \\
\(\chi_i\) & 50.0 & 70.0 & Negative curvature of \(A_i(m_i)\). \\
\(d_i\) (war) & 0.01 & 0.03 & Additional capital destruction in war. \\ \bottomrule
\end{tabular}
\end{table}

The United States is modeled as a high-productivity, high-savings economy with a
relatively strong capacity to convert defense-related research into broader
technological advances (high \(A_{0,U}\), \(s_U\), and \(\varphi_U\), and a relatively
low \(\chi_U\)). Iran is modeled as a lower-productivity economy with somewhat lower
savings, higher depreciation, and a less efficient and more distortionary
military--innovation channel (lower \(A_{0,I}\), \(s_I\), and \(\varphi_I\), higher
\(\chi_I\), and higher war-related destruction \(d_I\)). These choices are stylized,
but they reflect the empirical observation that advanced economies tend to have deeper
capital markets, more diversified innovation systems, and more resilient production
structures.

\subsection{Peace and War Scenarios}

To make the model concrete, we consider two regimes for each country: a peacetime
regime with moderate, time-invariant military burdens and no additional destruction,
and a wartime regime with sharply higher military burdens and positive destruction
rates. For the United States, we set a peacetime military burden of
\(m_U^{P} = 0.035\) (3.5\% of GDP), consistent with typical post-Cold War values, and a
wartime burden of \(m_U^{W} = 0.07\) (7\% of GDP) during a major conflict. For Iran, we
set a peacetime military burden of \(m_I^{P} = 0.03\) (3\% of GDP) and a wartime burden
of \(m_I^{W} = 0.10\) (10\% of GDP), reflecting the idea that an all-out war forces a
much larger share of national income into military uses in a smaller, more vulnerable
economy.

Using Equation~\eqref{eq:gi}, the implied long-run growth rates in each regime are
obtained by substituting the relevant parameter values and setting \(d_i = 0\) in
peacetime and \(d_i > 0\) in wartime. Table~\ref{tab:growth_war} summarizes the
results, expressing growth rates in percentage points for ease of interpretation.

\begin{table}[h!]
\centering
\caption{Simulated long-run growth under peace and war}
\label{tab:growth_war}
\begin{tabular}{@{}lcc@{}}
\toprule
Country & Peacetime growth \(g_i(m_i^{P},0)\) & Wartime growth \(g_i(m_i^{W},d_i)\) \\ \midrule
United States (\(U\)) & \(\approx 3.5\%\) & \(\approx 2.4\%\) \\
Iran (\(I\))          & \(\approx -1.4\%\) & \(\approx -6.2\%\) \\ \bottomrule
\end{tabular}
\end{table}

For the United States, moving from peace to war reduces the long-run growth rate from
around 3.5\% to 2.4\%. The advanced economy can sustain a positive growth rate even
under higher militarization and some destruction, because high baseline productivity,
a strong innovation system, and robust savings partially offset the negative effects.
For Iran, by contrast, even the peacetime configuration in this calibration delivers a
slightly negative long-run growth rate, reflecting structural stagnation under
sanctions and limited productivity. War then deepens the contraction dramatically, with
growth falling to around \(-6\%\). The combination of very high military burdens,
elevated depreciation, and substantial destruction leaves little room for civilian
capital accumulation and constrains the capacity to exploit any positive innovation
effects of militarization.

\subsection{Growth--Militarization Profiles by Country}

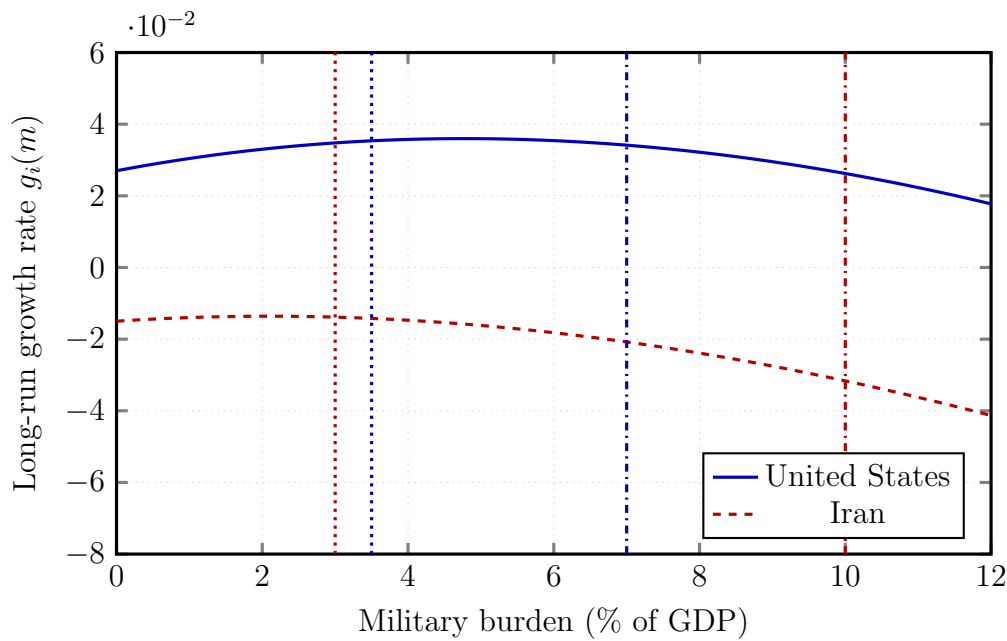
\begin{figure}[h!]
  \centering
  \begin{tikzpicture}
  \begin{axis}[
  width=0.8\textwidth,
  height=0.5\textwidth,
  xlabel={Military burden (\% of GDP)},
  ylabel={Long-run growth rate \(g_i(m)\)},
  xmin=0, xmax=0.12,
  ymin=-0.08, ymax=0.06,
  xtick={0,0.02,0.04,0.06,0.08,0.10,0.12},
  xticklabels={0,2,4,6,8,10,12},
  grid=both,
  grid style={dotted,gray!40},
  thick,
  axis line style={very thick},
  tick style={very thick},
  legend style={at={(0.97,0.03)},anchor=south east},
]
      % United States growth function
      \addplot[
        blue!75!black,
        domain=0:0.12,
        samples=400,
        very thick
      ]
      { -0.05 + 0.22*(1 - x)*0.35*(1 + 6*x - 50*x^2) };
      \addlegendentry{United States}

      % Iran growth function
      \addplot[
        red!70!black,
        domain=0:0.12,
        samples=400,
        very thick,
        dashed
      ]
      { -0.06 + 0.18*(1 - x)*0.25*(1 + 4*x - 70*x^2) };
      \addlegendentry{Iran}

      % Vertical lines for peacetime and wartime burdens (US)
      \addplot[blue!75!black, dotted, very thick] coordinates {(0.035,-0.08) (0.035,0.06)};
      \addplot[blue!75!black, dashdotted, very thick] coordinates {(0.07,-0.08) (0.07,0.06)};

      % Vertical lines for peacetime and wartime burdens (Iran)
      \addplot[red!70!black, dotted, very thick] coordinates {(0.03,-0.08) (0.03,0.06)};
      \addplot[red!70!black, dashdotted, very thick] coordinates {(0.10,-0.08) (0.10,0.06)};
    \end{axis}
  \end{tikzpicture}
  \caption{Simulated growth--militarization profiles for the United States and Iran.
  Note: theoretical simulation based on Equation~\eqref{eq:gi} with \(d_i = 0\). Vertical
  lines mark the calibrated peacetime (dotted) and wartime (dash-dotted) burdens.}
  \label{fig:g_two_country}
\end{figure}

It is instructive to examine how the long-run growth rate varies with the military
burden for each country, holding destruction rates at their peacetime values
(\(d_i = 0\)). Using Equation~\eqref{eq:gi} with \(d_i = 0\), we have
\begin{equation}
  g_U(m) = -0.05 + 0.22 (1 - m) \, 0.35 \bigl[1 + 6 m - 50 m^2\bigr],
\end{equation}
\begin{equation}
  g_I(m) = -0.06 + 0.18 (1 - m) \, 0.25 \bigl[1 + 4 m - 70 m^2\bigr],
\end{equation}
for the United States and Iran respectively. Figure~\ref{fig:g_two_country} plots these
functions over a range of military burdens from 0 to 12\% of GDP.

Both curves are hump-shaped, but their heights and locations differ. For the United
States, the maximum of the growth function occurs at a moderate military burden, and
the entire relevant range (up to about 8\% of GDP) includes positive growth rates,
reflecting strong underlying productivity and savings. For Iran, the entire curve lies
much lower, and the interval of military burdens that yields non-negative growth is
narrow and located at relatively low values of \(m\). High militarization quickly
pushes the economy into negative growth. This visualization captures the idea that the
same level of military burden can be macroeconomically sustainable for an advanced
economy but destructive for a structurally weaker one.

\subsection{Interpreting the Macroeconomic Costs of War}

Within this theoretical setup, war is represented by two changes: an increase in the
military burden and a rise in the effective depreciation rate due to destruction. For
the United States, the shift from \(m_U^{P} = 0.035, d_U = 0\) to
\(m_U^{W} = 0.07, d_U = 0.01\) moves the economy leftward along a relatively flat
portion of the hump-shaped curve and adds a small downward shift due to destruction.
The result is a reduction in the long-run growth rate of around one percentage point.
Over a decade, this difference compounds into a substantial loss of output relative to
the counterfactual peace path, but the economy continues to grow, and its structural
capacity is not fundamentally undermined.

For Iran, the move from \(m_I^{P} = 0.03, d_I = 0\) to \(m_I^{W} = 0.10, d_I = 0.03\)
is far more damaging. The increase in the military burden pushes the economy onto the
steeply downward-sloping segment of its growth curve, where additional militarization
sharply reduces the effective investment share and the net contribution of
military-induced innovation. The added destruction term then shifts the entire curve
downward. Together, these changes produce a collapse in the long-run growth rate from
approximately \(-1.4\%\) to \(-6.2\%\). In macroeconomic terms, the war transforms a
stagnating economy into one experiencing rapid cumulative contraction, with severe
implications for living standards and the capacity to rebuild after the conflict.

The asymmetry in outcomes is not merely a function of initial income levels, but of the
interaction between structural parameters: savings behavior, baseline productivity, the
shape of the military--innovation channel, and the vulnerability of the capital stock
to destruction. The model thus provides a simple analytical argument for why wars
between highly unequal economies tend to impose much heavier relative macroeconomic
costs on the weaker side, even when the stronger side also invests heavily in military
operations.

\section{Conclusion}

This paper has developed a theory-based framework for analyzing the economic
consequences of militarism and war. Starting from a simple Keynesian
income-expenditure model in which military spending operates as a component of
aggregate demand, it has extended the analysis into a dynamic endogenous-growth
setting where the composition of government expenditure between civilian and military
uses affects both capital accumulation and productivity growth. The central theoretical
result is a non-linear relationship between the military burden and the long-run growth
rate, generated by the interaction of crowding-out effects and defense-induced
innovation.

Using calibrated parameters inspired by standard macroeconomic magnitudes, the paper
has simulated this growth function and shown that, under plausible conditions, there
exists an interior military burden that maximizes long-run growth. Below this level,
the potential technological benefits of militarization are not fully exploited; above
it, excessive military spending depresses growth by reducing civilian investment and
distorting the direction of technical change. The concept of the permanent war economy
can be interpreted as a situation in which political and institutional forces keep the
economy on the downward-sloping segment of this growth function, locking in a pattern
of high militarization and suboptimal growth.

Extending the model to a two-country environment, with parameters chosen to reflect
stylized differences between the United States and Iran, has highlighted the
asymmetric macroeconomic costs of war. An advanced, diversified economy can sustain
positive growth under substantial militarization and limited destruction, although at a
significant opportunity cost relative to peace. A structurally weaker economy with
lower productivity, constrained savings, and higher vulnerability to destruction faces
a much steeper trade-off: war can easily push it into deep and persistent contraction.

Future research could extend the model in several directions. Endogenizing the military
burden through political economy mechanisms would allow a more complete analysis of how
militarized equilibria emerge and persist. Incorporating heterogeneous sectors and
explicit defense R\&D decisions would make it possible to study the distributional and
technological consequences of militarism in greater detail. Finally, empirical work
using panel data and structural estimation techniques could be undertaken to assess the
quantitative relevance of the theoretical mechanisms identified here.

\section*{Declerations}

This manuscript benefited from the use of ChatGPT (version 5.1) for technical writing support. All ideas, analysis, and interpretations are the responsibility of the author, who has thoroughly reviewed and revised the manuscript.

The author does not support any form of violence or extreme ideologies. This research also does not account for the human cost of war, particularly the loss of lives, as such considerations are beyond the scope of the present paper. The study, however, can be extended in several ways by incorporating real-world data and contemporary newspaper reports on the war, which represents a clear direction for future research.

\newpage

\bibliographystyle{apalike}

\end{document}